\pdfoutput=1

\documentclass[11pt]{article}

\usepackage[]{EMNLP2023}

\usepackage{times}
\usepackage{latexsym}

\usepackage[T1]{fontenc}

\usepackage[utf8]{inputenc}

\usepackage{microtype}

\usepackage{inconsolata}

\usepackage{graphicx}

\usepackage{multirow}
\usepackage{booktabs}
\usepackage{pdfpages}
\PassOptionsToPackage{hyphens}{url}\usepackage{hyperref}

\def\addingon{\texttt{Adding on}}
\def\eliciting{\texttt{Eliciting}}
\def\revoicing{\texttt{Revoicing}}
\def\probing{\texttt{Probing}}
\def\connecting{\texttt{Connecting}}
\def\poortranscription{\texttt{Poor transcription quality}}
\def\offtask{\texttt{Off task}}

\title{Measuring Five Accountable Talk Moves to Improve Instruction at Scale}
\author{Ashlee Kupor \\
  Stanford University \\
  \texttt{akupor@stanford.edu} \\\And
  Candice Morgan \\
  Stanford University \\
  \texttt{cpenelto@stanford.edu} \\ \And
  Dorottya Demszky \\
  Stanford University \\
  \texttt{ddemszky@stanford.edu} \\}

\begin{document}
\maketitle

\begin{abstract}
Providing consistent, individualized feedback to teachers on their instruction can improve student learning outcomes. Such feedback can especially benefit novice instructors who teach on online platforms and have limited access to instructional training. To build scalable measures of instruction, we fine-tune RoBERTa and GPT models to identify five instructional talk moves inspired by accountable talk theory: \addingon{}, \connecting{}, \eliciting{}, \probing{} and \revoicing{} students' ideas. We fine-tune these models on a newly annotated dataset of 2500 instructor utterances derived from transcripts of small group instruction in an online computer science course, Code in Place. Although we find that GPT-3 consistently outperforms RoBERTa in terms of precision, its recall varies significantly. We correlate the instructors' use of each talk move with indicators of student engagement and satisfaction, including students' section attendance, section ratings, and assignment completion rates. We find that using talk moves generally correlates positively with student outcomes, and connecting student ideas has the largest positive impact. These results corroborate previous research on the effectiveness of accountable talk moves and provide exciting avenues for using these models to provide instructors with useful, scalable feedback.\footnote{Models and scripts can be found at: \url{https://huggingface.co/aekupor/talk-move-router.}}%
\end{abstract}

\section{Introduction}
\label{sec:intro}


Online education is growing at an unprecedented speed, partially fueled by the lasting impact of the COVID-19 pandemic. 
Among online courses, those that offer live instructor-student interaction are the most beneficial for student learning \citep{sun_teacher_2022}. To scale small group or 1:1 instruction, many platforms recruit novice instructors who receive limited pedagogical training. Recent work has shown that providing novice instructors with automated feedback can improve their instruction and student learning outcomes in a cost-effective and scalable way in a 1:1 context \citep{demszky2023mpowering} and in a small group setting as part of a free online programming course, Code in Place \citep{demszky2023can}. The feedback in Code in Place was powered by a model that identified a single talk move: instructor's uptake of student ideas and was fine-tuned on transcripts of math instruction rather than computer science \citep{demszky2021measuring}. Creating models that are fine-tuned to identify a wider set of pedagogical talk moves and are specifically adapted to computer science data can enhance feedback that online platforms can provide to computer science instructors.

\begin{figure}
    \centering
    \includegraphics[width=\linewidth]{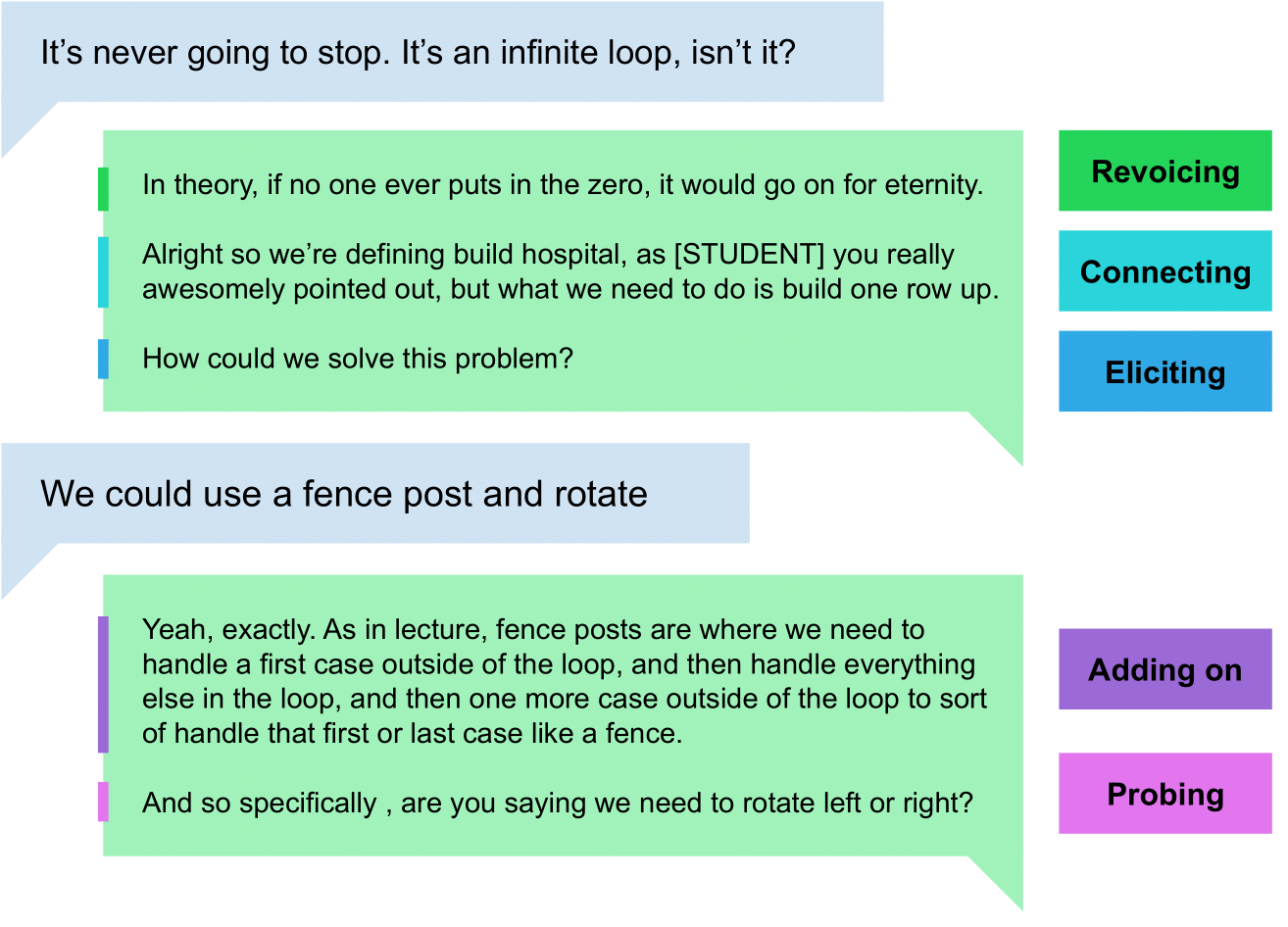}
    \caption{Example teacher utterances demonstrating the 5 talk moves.}
    \label{fig:my_label}
\end{figure}

We develop five open-source models, using transcripts from Code in Place 2021, that measure talk moves inspired by the Accountable Talk framework \citep{connor_2015_scaling}: \addingon{}, \connecting{}, \eliciting{}, \probing{} and \revoicing{} students' ideas. This framework has been used in the past to develop models and provide automated feedback to K-12 math teachers \citep{suresh_using_ai_2021,jacobs_promoting_2022}. Additionally, Code in Place 2023 decided to create training for these talk moves based on input from faculty in education, Section Leaders of the introductory Computer Sciences classes at Stanford University, other departments to determine the best practices for instructors to encourage student learning and engagement. Thus, the choice to use Accountable Talk Theory was inspired by a long line of work in theory as well as knowledge and expertise of people who were involved in the course.

We recruit computer science instructors to create an annotation guide and annotate 2500 instructor utterances for these five talk moves. We fine-tune RoBERTa \citep{liu2019roberta} and OpenAI's GPT-3 \citep{brown2020language} on the annotated data. We find that GPT-3's precision is consistently higher than RoBERTa, but that its recall varies significantly across talk moves.

We run inference over all four thousand transcripts from Code in Place 2021 to predict talk moves, and correlate the frequency of talk moves with indicators of student engagement and satisfaction, including students' section attendance, section ratings, and assignment completion rates. We find that using talk moves generally correlates positively with student outcomes, with \connecting{} student ideas showing the most positive impact. These findings corroborate previous research on the effectiveness of talk moves and show promise for using these models to give computer science instructors useful, personalized, and scalable feedback to improve their teaching.

\section{Related Work}
\label{sec:related_work}




Traditional methods of instructor feedback are slow and costly. Advances in natural language processing have created avenues for quick, scalable, and useful feedback for instructors. Researchers have measured question authenticity (\citealt{Cook2018AnOV}; \citealt{kelly_2018}) and have worked to distinguish questions from live classroom audio (\citealt{donnelly_2017}; \citealt{blanchard_2016}). Researchers have used machine learning to score different dimensions of instructor discourse \citep{jensen_2022} and to classify semantics of classroom dialogue \citep{song_automatic}. 

The talk moves we chose are part of the Accountable Talk Framework that encourages instructors to promote student participation in an equitable way \citep{connor_2015_scaling}. These talk moves have been found to facilitate discussions in ways that promote learning (\citealp{Michaels2007-MICDDI}; \citealp{jacobs_promoting_2022}) and can help promote equity in the classroom \citep{OCONNOR2019166}. The talk move framework provides examples for instructors to build community, content knowledge, and promote rigorous thinking (\citealp{suresh-etal-2022-talkmoves}; \citealp{suresh_using_ai_2021}). \citet{ganesh2021teacher} has used neural networks to predict the next talk move that a instructor will use. \citet{demszky2021measuring} used an unsupervised approach to measure instructors' uptake of student contributions.

In the last few years, researchers have begun to compare performance between RoBERTa and GPT, but not yet in terms of classification tasks as we do. Using logical reasoning benchmarks, ChatGPT performs significantly better than RoBERTa, and GPT-4 performs even better \citep{liu2023evaluating}. Additionally, for few-shot prompting, GPT-3 is found to improve upon RoBERTa \citep{si2023prompting}. However, in a biomedical context for information extraction, GPT-3 performs significantly worse than RoBERTa \citep{gutiérrez2022thinking}.

\section{Data}
To generate our sample for annotation, we (i) transcribed all $\sim$4k small group instruction recordings via OpenAI's Whisper \citep{radford_robust_2022}, (ii) diarized the transcripts using speaker timestamps from the video recording and (iii) merged the transcripts derived from the audio with chat logs to additionally account for written interaction. Last, (iv) we randomly sampled 2000 instructor utterances from these transcripts, (v) splitting any utterance over 200 tokens into multiple segments that were annotated in sequence. This process resulted in $2,503$ unique annotation examples.


\paragraph{Annotation guide.} We developed the annotation guide consulting related work \citep{Michaels2007-MICDDI,jacobs_promoting_2022} and six undergraduate instructors with training and experience as section leaders in an introductory programming course covering the same material as Code in Place. We recruited these section leaders as annotators and conducted multiple rounds of practice sessions to empirically test and fine-tune our guide. In addition to the five talk moves, annotators also labelled teacher utterances for \offtask{} instruction (not related to programming) and indicate \poortranscription{} when the transcription quality interfered with labeling talk moves. The guide is included among the Supplementary Materials.

\begin{table}[]
\small
\resizebox{\linewidth}{!}{%
\begin{tabular}{ll}
\toprule
\textbf{Label} & \textbf{Description}  \\ \midrule



\begin{tabular}[l]{@{}l@{}}Adding On\\ (25\%) \end{tabular}  & Utterance builds on students' ideas.  \\ \midrule

\begin{tabular}[l]{@{}l@{}}Connecting\\ (4\%) \end{tabular} & \begin{tabular}[l]{@{}l@{}}Utterance incorporates students's \\prior ideas into current discussion.\end{tabular} \\ \midrule

\begin{tabular}[l]{@{}l@{}}Eliciting\\ (17\%) \end{tabular} & \begin{tabular}[l]{@{}l@{}}Utterance encourages students \\ to initiate problem solving \\ and share ideas.\end{tabular} \\ \midrule

\begin{tabular}[l]{@{}l@{}}Probing\\ (13\%) \end{tabular} & \begin{tabular}[l]{@{}l@{}}Utterance gets students to elaborate \\ and clarify their expressed ideas.\end{tabular} \\ \midrule

\begin{tabular}[l]{@{}l@{}}Revoicing\\ (11\%) \end{tabular} & \begin{tabular}[l]{@{}l@{}}Utterance restates students’ ideas,\\  anticipating students’ intent.\end{tabular}  \\ 


\bottomrule
\end{tabular}%

}
\caption{Talk move definitions with label distribution in parentheses.}
\label{tab:talk_move_desc}
\end{table}


\paragraph{Talk move distribution.} Two randomly selected annotators labelled each utterance, and for 63\% of examples, they made the same selection for at least one label. 
To evaluate annotator agreement, for each rater and talk move, we calculate how often their judgment agrees with the other rater. Then, we compute the average of the six raters' agreement scores for each talk move (\addingon{}: 81\%, \connecting{}: 97\%, \eliciting{}: 90\%, \probing{}: 91\%, \revoicing{}: 93\%). 

Talk moves are sparse in the data, and annotators labelled the most common one, \addingon{}, 25\% of the time and the least frequent one, \connecting{}, only 4\% of the time (Table~\ref{tab:talk_move_desc}). Due to their sparsity, we label an example with a talk move if either annotator selected that label. Only 4\% of talk moves were marked as poor quality and we had a WER score of 12\%.

\section{Classifying Talk Moves}

\paragraph{Methods.} We fine-tuned separate models for each of the five talk moves. We experimented with various preprocessing options: (i) including the two prior utterances before the instructor utterance to provide context
, (ii) if the prior text was included, truncating the prior text either from the beginning or end to ensure we didn't go over the token limit, and (iii) balancing binary labels using different ratios (ranging between 1:1 to 1:8, representing the ratio of 1 to 0 labels) to mitigate sparsity for certain talk moves. See Appendix \ref{sec:model_hyperparameters} for details. 

We identified the best data preprocessing options using RoBERTa (Appendix \ref{sec:best_RoBERTa_models}) and used the same data to fine-tune GPT-3 with some additional processing to expected data format (Appendix \ref{sec:best_gpt_models}). We fine-tuned models on the same training split (80\%) and evaluated them on the same held-out set (20\%). 

A singular utterance can be marked as multiple talk moves. Each model that labels an utterance as a talk move is independent of the others, so an utterance can be labeled as an example of multiple talk moves. Each model is a binary classification, so that we can examine utterances for just a singular talk move, or we could look at combinations of different talk moves together.

\paragraph{Results.} Our results are displayed in Table \ref{tab:comparing_models}. In terms of F1 scores, we find that GPT-3 typically only performs a few points better than RoBERTa for most talk moves, and performs significantly worse for \connecting{}, as we discuss later. The F1 scores range between .5-.75, choosing the best model for each talk move, due to data sparsity and the complexity of the task.

In terms of precision, GPT-3 has a higher score for every talk move, being at least 5 points higher and sometimes as much as 21 points more. We see the least improvement for precision for \eliciting{}, but RoBERTa's precision for \eliciting{} was already the highest of all talk moves to begin with. We see the largest improvement of 21 points for \probing{}. This improvement likely comes from GPT being able to more accurately tell the different between \eliciting{} and \probing{}.

\begin{table}\centering
\footnotesize
\resizebox{\linewidth}{!}{%
\begin{tabular}{lccc|cccc}\toprule
\textbf{Model} &\multicolumn{3}{c}{\textbf{RoBERTa}} &\multicolumn{3}{c}{\textbf{GPT-3}} \\

&Precision &Recall &F1 &Precision &Recall &F1 \\\midrule 

Adding on &0.61 &0.70 &0.65 &0.77 &0.62 &0.68 \\
Connecting &0.40 &0.67 &0.50 &0.50 &0.20 &0.29 \\
Eliciting &0.72 &0.63 &0.67 &0.77 &0.73 &0.75 \\
Probing &0.39 &0.56 &0.46 &0.60 &0.56 &0.58 \\
Revoicing &0.45 &0.45 &0.45 &0.56 &0.45 &0.50 \\
\bottomrule
\end{tabular} %
}
\caption{
Comparing RoBERTa and GPT-3 precision, recall, and F1 scores for each talk move model.
}
\label{tab:comparing_models}
\end{table}

In terms of recall, GPT-3 varies compared to RoBERTa: it performs the same as RoBERTa for \probing{} and \revoicing{}, 10 points better for \eliciting{}, and up to 47 points worse for \connecting{}. We see GPT-3's lowest recall scores for \connecting{}, most likely because \connecting{} has the fewest training examples, and using a low balancing factor may have led GPT to overfit. 

GPT-3 F1's score is better for all talk moves with improvements ranging between 3-12 points compared to RoBERTa, except for \connecting{} where it performs over 20 points worse. \connecting{} has the fewest training examples, and probably does not have enough examples to adequately train a GPT-3 model. In future iterations, we hope to annotate more data so that we will have more \connecting{} training examples to be able to properly train and compare the models. Ultimately, GPT-3 does typically perform better than RoBERTa in terms of F1 scores, but the size of difference depends on the specific talk move. Given that GPT-3 is expensive and only improves a little over the RoBERTa models, we use RoBERTa models for inference on the entire dataset in the next section.

\section{Correlation With Student Outcomes}
\label{sec:student_outcomes}

\begin{table}[]
\centering
\footnotesize
\begin{tabular}{@{}lccc@{}}
\toprule
  \begin{tabular}[l]{@{}l@{}}Talk Move\\ (Independent\\Variable)\end{tabular}        & \textbf{\begin{tabular}[c]{@{}c@{}}Subsequent\\Section\\ Attendance\end{tabular}} & \textbf{\begin{tabular}[c]{@{}c@{}}Section\\ Helpfulness\\ Rating\end{tabular}} & \textbf{\begin{tabular}[c]{@{}c@{}}Number of\\ Completed\\ Assignments\end{tabular}} \\ \midrule
Adding On  & -0.012**                                                                         & 0.010**                                                                       & -0.006**                                                                      \\
           Connecting & 0.067**                                                                          & 0.038*                                                                        & 0.034**                                                                       \\ 
Eliciting  & 0.007*                                                                           & 0.012**                                                                       & 0.002                                                                         \\
Probing    & -0.002                                                                           & 0.008*                                                                        & 0.000                                                                         \\

Revoicing  & -0.003                                                                           & 0.010                                                                         & -0.004                                                                        \\

\bottomrule
\end{tabular}
\caption{$p$-values indicate significance: * <0.05, **<0.01. Coefficients show the correlation between the hourly rate of each talk move, measured by RoBERTa. Each cell represents a separate linear model, each of which contain the same student and instructor demographic control variables (Section~\ref{sec:student_outcomes}).}
\label{tab:regression_results}
\end{table}

Our ultimate goal is to improve student learning by providing feedback to instructors on talk moves. Here we conduct analyses to understand if an increased use of these talk moves correlates with student outcomes in our domain. To study this question, we (i) take the best RoBERTa classifiers for each talk move, (ii) predict talk moves for the entire dataset ($\sim$4k transcripts), and (iii) run regressions using the key student outcomes available in this dataset: attendance, students' section ratings and assignment completions. Student attrition was a major challenge in this optional online course, hence each of these three variables are used as important indicators of student engagement and satisfaction. 

\paragraph{Model.} We run linear regressions, clustering standard errors at the instructor level.  The models are captured by the equation: $
    y_d = x_m\beta_1 + T\beta_2 + S\beta_3
+ \varepsilon $, 
where $y_d$ is a vector representing a dependent variable $d$, $x_m$ is a vector representing the hourly rate of a talk move $m$ in a transcript, $T$ is a matrix of instructor covariates, $S$ is a matrix of student covariates, $\beta_1,\beta_2,\beta_3$ are vectors of parameters to be estimated and $\epsilon$ is a vector of residuals. 

The values reported are beta coefficients which indicate the relationship between the predictor (i.e. frequency of a talk move), and the outcome (e.g. attendance), controlling for covariates such as instructor and classroom demographics (discussed later). For example, a coefficient of 0.067 for connecting (predictor) and attendance (outcome) means that a use of 1 additional connecting move increases student attendance in the subsequent section by 0.067 students. We multiply by 10 to improve interpretability so that 10 additional connecting moves increase student attendance in the subsequent section by close to one student on average. 

\paragraph{Dependent variables.} For the first model, we use the number of students attending the subsequent section as a dependent variable, and we run analyses at the transcript level (N=2895). For the second model, we use ratings of section helpfulness as reported in the student exit survey on a four point scale, and conduct analyses at the student level (N=774). For the third model, we use the number of optional assignments completed (out of 3), also conducting analyses at the student level (N=909).

\paragraph{Covariates.} We include several covariates for instructor and classroom demographics. We include an indicator for the instructor being female, whether they are first-time Code in Place instructors, whether they are located in the U.S. and variables for their age (age and age$^2$, following standard practice in education research). We include variables related to section composition in terms of proportion of female students, students in the U.S. and students in different age buckets (22-25, 26-30, 31-35, 36-40, 40+, as collected by the course). 

\paragraph{Results.} The results are summarized in Table~\ref{tab:regression_results}. Two key patterns emerge: \connecting{}, the least frequent talk move correlates most positively with all of the outcomes. For example, using 10 more connecting moves in an hour-long session results increases section attendance by .67 (30\% more than the average), course ratings by .38 (17\%) and number of completed assignments by .34 (34\%). This indicates that \connecting{} student ideas may be the most challenging to adopt but also the most high-leverage talk move for improving instruction quality. In contrast, \addingon{}, the most frequent talk move correlates negatively with attendance and assignment completions, perhaps suggesting that this talk move correlates with the section leader (over-)explaining concepts rather than encouraging student participation. Interestingly, students still perceive \addingon{} as a positive practice as indicated by its positive correlation with section ratings. Questioning strategies also show a positive impact, with \eliciting{} and \probing{} both correlating positively with section ratings and \eliciting{} also correlating positively with student attendance. Interestingly, \revoicing{} is the only talk move that does not show a significant correlation with either outcome, despite the fact that uptake---a closely related move---was found to correlate positive with student outcomes~\citep{demszky2021measuring}, indicating differences between the two measures that are worth further investigating.

\section{Conclusion}

We develop RoBERTa and GPT-3 models to detect five accountable talk moves in instructor utterances, training on a newly annotated dataset of transripts derived from online, small-group computer science instruction. Using these models, we find that the use of talk moves is generally positively correlated with student outcomes, with \connecting{} having the most positive impact, followed by \eliciting{}. These models provide promising avenues for giving instructors feedback, especially for those who have little to no teaching experience and teach classes online. We hope to do further research using these models to analyze how feedback can best be given to instructors in an effective and helpful manner so that they can easily synthesize recommendations and improve their teaching skills.

\section*{Limitations}
\label{sec:limitations_future_work}

We relied on human annotators to label the data, which is costly, and thus we only had about 2500 labeled examples. With additional training data, our models could be more accurate. Additionally, the annotators do not have perfect agreement, which can affect the accuracy of our models.

The training data was from an introductory computer science class, and hasn't been verified on other subjects. The data is also all in English, and hasn't been verified on other languages. In future iterations, we hope to diversify our training data to improve generalizability.

As RoBERTa and GPT both have limits on the number of tokens, we had to limit the length of each utterance. In cases where the instructor spoke for a while, we split up their utterance into multiple smaller utterances, potentially splitting the use of a talk move into different utterances. Although the instructor did use a talk move, our model may not label it as such since it was split into different utterances that are each analyzed individually by the model. Thus, there are instances where talk moves would have been found if the text wasn't split at that instance. In future iterations, we hope to utilize a sliding window approach or another method to avoid splitting up talk moves.

In addition, the model currently labels an entire utterance as either an example of the talk move or not. In the future, we hope to have the model highlight the specific part of the utterance where the talk move is. For example, the model currently would label a 3 sentence utterance as \eliciting{}, when the actual \eliciting{} may only be 1 sentence. Future iterations of the models would return only the specific phrase or phrases where the talk move is used, rather than the entire utterance.

\section*{Ethics Statement}
\label{sec:ethics}

The training data represents instructor utterances from a certain population that may not generalize to other populations. The training data is taken from transcripts of instructors who taught sections of an online introductory computer science course in English. Almost 70\% of the instructors are male, with almost 65\% from North America. The average age was about 30 years old and almost 80\% of the instructors were first time instructors. The transcripts obtained from these instructors represent a certain population, so it requires careful validation and testing to see if the models can generalize to other populations. Additionally, the data was evaluated by 6 annotators, who, even though they come from diverse racial and gender backgrounds, come from the same undergraduate institution and are all college aged, which may also lead to biases in how the training data was annotated.

These models have promising avenues for instructor professional learning. We have an ongoing study that uses these models to deliver feedback to instructors weekly over the period of a course. We see a great opportunity in using these models to give instructors specific, timely, scalable, and low-cost feedback to improve on their teaching, but we must be careful of potential side effects from this feedback. The model does not have perfect accuracy, and presenting instructors with potentially incorrect feedback can affect not only their teaching performance, but their self-confidence in their teaching. Given that these models have some false positives and negatives, they should be used carefully when providing feedback to instructors. It is essential to study the differential impact that the feedback can have on instructors and to understand how feedback on these different talk moves can best be integrated into instructor professional development. 

We believe that these models should complement, and not replace, other forms of professional development for instructors. These models are just a tool that can be used by instructors, and are not meant to replace expert training and feedback. These models should not be used to profile instructors or to make high-stakes decisions about hiring or firing instructors.

\section*{Acknowledgements}
This work was supported by a grant from the Schmidt Futures Foundation. We are grateful to Thanawan (Ly-Ly) Atchariyachanvanit, Bryant Perkins, Iddah Ashley Kudakwashe Mlauzi, Ryan Guan for help with data annotation. We thank Abhinav Garg for help with transcribing the data.

\bibliography{anthology,custom}
\bibliographystyle{acl_natbib}

\appendix

\section{Data Preprocessing}
\label{sec:model_hyperparameters}

For each RoBERTa model, we experimented with the following different data preprocessing options:

\paragraph{Including prior texts.} For some talk moves, specifically \addingon{} and \probing{}, it is intuitive that the model would need to see utterances prior to just the instructor utterance to make an accurate classification. For example, it would be extremely difficult for the model to tell when the instructor was \addingon{} to something a student just said if it is unable to see the prior student utterance. 


For long instructor utterances, the utterances were split into smaller segments, but the prior text was the same for all of the smaller utterances. 

The maximum sentence length of RoBERTa is 512 tokens, so if the prior text plus the instructor utterance was longer than 512, we needed to cut the prior text in some way. We experimented with two different ways of truncating the prior text:

\emph{Truncate from beginning.} In this variation, we would take the first part of the prior text until we reached the token limit. In practice, this means that a portion of the prior utterance right before the instructor utterance was cut off, and that the prior text consisted of mostly of the utterance two previous of the instructor.

\emph{Truncate from end.} Instead of cutting off the end of the prior text when the token limit was reached, and losing the text that was said right before the instructor utterance, another variation was that we would cut the beginning of the prior text. We included most of prior utterance right before the instructor utterance until the token limit was met. In practice, this means that the prior text consisted mostly of the utterance directly before the instructor utterance.

\paragraph{Balancing labels.} For some talk moves, especially for \connecting{}, they were not used frequently in the transcripts and thus we did not have many training examples. We experimented with balancing the labels with different balancing factors. To do this, we found the number of examples in our training dataset that didn’t exhibit the talk move, and then divided this number by the balancing factor to find the number of training examples we needed to have that did exhibit this talk move. If there weren’t enough training examples to hit that number, we would continue to sample examples exhibiting that talk move until we hit that number (so some examples were included multiple times). Intuitively, this gives the model a larger set of examples exhibiting a talk move, so that it can make more accurate predictions. However, if the balancing was too high, the model could become over-fit to the training data and not generalize well to new utterances.

\section{Best RoBERTa models}
\label{sec:best_RoBERTa_models}

We tried the many combinations of data preprocessing (including prior text, where to truncate the prior text, balancing labels) for each of the RoBERTa models. We found the following to be the best models:

\paragraph{Adding On:} two prior texts with truncating from the beginning, no balancing. \addingon{} consists of adding additional information to what students have already said, and thus the prior text is needed. We found the best model to be when we take the beginning of the two prior texts rather than the last part, as we do for \probing{} and \revoicing{}. For example, an instructor might be discussing a topic when a student asks a clarifying question and the instructor then adds on to what they were explaining, and thus we want to include more of what the instructor was saying two utterances prior to the utterance being examined to understand the context. There were many examples of this so we did not need to use balancing. 

\paragraph{Connecting:} no prior text, balancing with balancing factor of 6. \connecting{} is typically done by mentioning another student’s thought or idea and then asking students to comment off of it, and does not need the prior text to effectively determine an utterance. \connecting{} is the least common talk move, so there was a high balancing factor in order to aptly train the model.

\paragraph{Eliciting:} no prior text, no balancing. \eliciting{} mostly consists of asking questions to start getting students to think about the problem at hand, and thus the prior text isn’t needed to determine an utterance. \eliciting{} was quite common, so there were plenty of examples to aptly train the model, and thus balancing was not needed.

\paragraph{Probing:} two prior texts with truncating from the end, no balancing. \probing{} involves asking deeper questions about student ideas, and thus the prior text is needed to determine what the student was discussing prior to the instructor probing. We found truncating from the end to be best as the instructor often will probe into what the student most recently said, and thus having the student’s last few phrases is helpful. There were many \probing{} examples and thus balancing was not needed.

\paragraph{Revoicing:} two prior texts with truncating from the end, balancing with balancing factor of 1. \revoicing{} includes repeating what the student just said, but in a different way, and thus the prior text is needed to see what the student said right beforehand. We used balancing with just a small balancing factor since \revoicing{} examples were quite sparse.

\section{Best GPT-3 models}
\label{sec:best_gpt_models}

To find the best GPT models, we experimented with different numbers of epochs using both the curie and davinci models. We start at a maximum of 5 epochs and then decrease the number of epochs. Once the best curie model is found, we used the same number of epochs with davinci.

Interestingly, \eliciting{} was the only model where davinci performed better than curie. However, the davinci model only improved 1\% over curie, so the two models perform practically the same. For the other talk moves, davinci performed slightly worse than the curie model, but was typically within a few points.

The best GPT models and their hyperparameters are:

\emph{Adding on:} curie model using 4 epochs.

\emph{Connecting:} curie model using 3 epochs. 

\emph{Eliciting:} davinci model using 5 epochs. 

\emph{Probing:} curie model using 4 epochs.

\emph{Revoicing:} curie model using 5 epochs.

\section{Model Utterances}

We also built a model to label model utterances, which can be any of the other 5 talk moves, but is an exemplary use case of one of them.

The best RoBERTa model we found was using two prior texts with truncating from the end and balancing with a balancing factor of 1. This model found model utterances for all the talk moves, and thus its data preprocessing reflect the most common preprocessing of the other talk move models.

The best GPT model we found was the curie model using 5 epochs.

We found RoBERTa's precision to be 0.29 compared to GPT's 0.59. This is the largest improvement we saw for any talk move. Model utterances have the widest variety of examples, as they can include any of the other 5 talk moves. Thus, GPT is able to understand this complexity and be more precise than RoBERTa. 

For recall, GPT performed worse at 0.23 compared to RoBERTa's 0.41. The F1 scores were nearly identical, with RoBERTa at 0.34 and GPT at 0.33.

\section{Annotation Guide}
\label{sec:anotation_guide}

The annotation guide we gave to all annotators can be found at: \url{https://docs.google.com/document/d/1UcDv80nW_j5ueP-2eyVSxHxWXp2BTkFDnXbFG_aXfiQ/edit?usp=sharing}. We used this guide to train the annotators before starting to annotate the training data.
 



\end{document}